\documentstyle[12pt]{article}

\catcode`\@=11
\long\def\@makefntext#1{ 
\protect\noindent \hbox to 3.2pt {\hskip-.9pt
$^{{\ninerm\@thefnmark}}$\hfil}#1\hfill} 

\def\thefootnote{\fnsymbol{footnote}}
 \def\@makefnmark{\hbox to 0pt{$^{\@thefnmark}$\hss}}  

\def\ps@myheadings{\let\@mkboth\@gobbletwo
\def\@oddhead{\hbox{} 
\rightmark\hfil\ninerm\thepage}
\def\@oddfoot{}\def\@evenhead{\ninerm\thepage\hfil 
\leftmark\hbox{}}\def\@evenfoot{}
\def\sectionmark##1{}\def\subsectionmark##1{}}

\begin{document}

\newcommand{\symbolfootnote}{\renewcommand{\thefootnote}
	{\fnsymbol{footnote}}}
\renewcommand{\thefootnote}{\fnsymbol{footnote}}
\newcommand{\alphfootnote}
	{\setcounter{footnote}{0}
	 \renewcommand{\thefootnote}{\sevenrm\alph{footnote}}}

\newcounter{sectionc}\newcounter{subsectionc}\newcounter{subsubsectionc}
\renewcommand{\section}[1] {\vspace{0.6cm}\addtocounter{sectionc}{1}
\setcounter{subsectionc}{0}\setcounter{subsubsectionc}{0}\noindent
	{\bf\thesectionc. #1}\par\vspace{0.4cm}}
\renewcommand{\subsection}[1] {\vspace{0.6cm}\addtocounter{subsectionc}{1}
	\setcounter{subsubsectionc}{0}\noindent
	{\it\thesectionc.\thesubsectionc. #1}\par\vspace{0.4cm}}
\renewcommand{\subsubsection}[1]
{\vspace{0.6cm}\addtocounter{subsubsectionc}{1}
	\noindent {\rm\thesectionc.\thesubsectionc.\thesubsubsectionc.
	#1}\par\vspace{0.4cm}}
\newcommand{\nonumsection}[1] {\vspace{0.6cm}\noindent{\bf #1}
	\par\vspace{0.4cm}}

\newcounter{appendixc}
\newcounter{subappendixc}[appendixc]
\newcounter{subsubappendixc}[subappendixc]
\renewcommand{\thesubappendixc}{\Alph{appendixc}.\arabic{subappendixc}}
\renewcommand{\thesubsubappendixc}
	{\Alph{appendixc}.\arabic{subappendixc}.\arabic{subsubappendixc}}

\renewcommand{\appendix}[1] {\vspace{0.6cm}
        \refstepcounter{appendixc}
        \setcounter{figure}{0}
        \setcounter{table}{0}
        \setcounter{equation}{0}
        \renewcommand{\thefigure}{\Alph{appendixc}.\arabic{figure}}
        \renewcommand{\thetable}{\Alph{appendixc}.\arabic{table}}
        \renewcommand{\theappendixc}{\Alph{appendixc}}
        \renewcommand{\theequation}{\Alph{appendixc}.\arabic{equation}}
        \noindent{\bf Appendix \theappendixc #1}\par\vspace{0.4cm}}
\newcommand{\subappendix}[1] {\vspace{0.6cm}
        \refstepcounter{subappendixc}
        \noindent{\bf Appendix \thesubappendixc. #1}\par\vspace{0.4cm}}
\newcommand{\subsubappendix}[1] {\vspace{0.6cm}
        \refstepcounter{subsubappendixc}
        \noindent{\it Appendix \thesubsubappendixc. #1}
	\par\vspace{0.4cm}}

\def\abstracts#1{{
	\centering{\begin{minipage}{30pc}\tenrm\baselineskip=12pt\noindent
	\centerline{\tenrm ABSTRACT}\vspace{0.3cm}
	\parindent=0pt #1
	\end{minipage} }\par}}

\newcommand{\bibit}{\it}
\newcommand{\bibbf}{\bf}
\renewenvironment{thebibliography}[1]
	{\begin{list}{\arabic{enumi}.}
	{\usecounter{enumi}\setlength{\parsep}{0pt}
\setlength{\leftmargin 1.25cm}{\rightmargin 0pt}
	 \setlength{\itemsep}{0pt} \settowidth
	{\labelwidth}{#1.}\sloppy}}{\end{list}}

\topsep=0in\parsep=0in\itemsep=0in
\parindent=1.5pc

\newcounter{itemlistc}
\newcounter{romanlistc}
\newcounter{alphlistc}
\newcounter{arabiclistc}
\newenvironment{itemlist}
    	{\setcounter{itemlistc}{0}
	 \begin{list}{$\bullet$}
	{\usecounter{itemlistc}
	 \setlength{\parsep}{0pt}
	 \setlength{\itemsep}{0pt}}}{\end{list}}

\newenvironment{romanlist}
	{\setcounter{romanlistc}{0}
	 \begin{list}{$($\roman{romanlistc}$)$}
	{\usecounter{romanlistc}
	 \setlength{\parsep}{0pt}
	 \setlength{\itemsep}{0pt}}}{\end{list}}

\newenvironment{alphlist}
	{\setcounter{alphlistc}{0}
	 \begin{list}{$($\alph{alphlistc}$)$}
	{\usecounter{alphlistc}
	 \setlength{\parsep}{0pt}
	 \setlength{\itemsep}{0pt}}}{\end{list}}

\newenvironment{arabiclist}
	{\setcounter{arabiclistc}{0}
	 \begin{list}{\arabic{arabiclistc}}
	{\usecounter{arabiclistc}
	 \setlength{\parsep}{0pt}
	 \setlength{\itemsep}{0pt}}}{\end{list}}

\newcommand{\fcaption}[1]{
        \refstepcounter{figure}
        \setbox\@tempboxa = \hbox{\tenrm Fig.~\thefigure. #1}
        \ifdim \wd\@tempboxa > 6in
           {\begin{center}
        \parbox{6in}{\tenrm\baselineskip=12pt Fig.~\thefigure. #1 }
            \end{center}}
        \else
             {\begin{center}
             {\tenrm Fig.~\thefigure. #1}
              \end{center}}
        \fi}

\newcommand{\tcaption}[1]{
        \refstepcounter{table}
        \setbox\@tempboxa = \hbox{\tenrm Table~\thetable. #1}
        \ifdim \wd\@tempboxa > 6in
           {\begin{center}
        \parbox{6in}{\tenrm\baselineskip=12pt Table~\thetable. #1 }
            \end{center}}
        \else
             {\begin{center}
             {\tenrm Table~\thetable. #1}
              \end{center}}
        \fi}

\def\@citex[#1]#2{\if@filesw\immediate\write\@auxout
	{\string\citation{#2}}\fi
\def\@citea{}\@cite{\@for\@citeb:=#2\do
	{\@citea\def\@citea{,}\@ifundefined
	{b@\@citeb}{{\bf ?}\@warning
	{Citation `\@citeb' on page \thepage \space undefined}}
	{\csname b@\@citeb\endcsname}}}{#1}}

\newif\if@cghi
\def\cite{\@cghitrue\@ifnextchar [{\@tempswatrue
	\@citex}{\@tempswafalse\@citex[]}}
\def\citelow{\@cghifalse\@ifnextchar [{\@tempswatrue
	\@citex}{\@tempswafalse\@citex[]}}
\def\@cite#1#2{{$\null^{#1}$\if@tempswa\typeout
	{IJCGA warning: optional citation argument
	ignored: `#2'} \fi}}
\newcommand{\citeup}{\cite}

\def\fnm#1{$^{\mbox{\scriptsize #1}}$}
\def\fnt#1#2{\footnotetext{\kern-.3em
	{$^{\mbox{\sevenrm #1}}$}{#2}}}

\font\twelvebf=cmbx10 scaled\magstep 1
\font\twelverm=cmr10 scaled\magstep 1
\font\twelveit=cmti10 scaled\magstep 1
\font\elevenbfit=cmbxti10 scaled\magstephalf
\font\elevenbf=cmbx10 scaled\magstephalf
\font\elevenrm=cmr10 scaled\magstephalf
\font\elevenit=cmti10 scaled\magstephalf
\font\bfit=cmbxti10
\font\tenbf=cmbx10
\font\tenrm=cmr10
\font\tenit=cmti10
\font\ninebf=cmbx9
\font\ninerm=cmr9
\font\nineit=cmti9
\font\eightbf=cmbx8
\font\eightrm=cmr8
\font\eightit=cmti8



\begin{flushright}RU-94-87
\end{flushright}

\centerline{\tenbf CP/CPT EXPERIMENTS WITH NEUTRAL KAONS}
\centerline{\tenbf OR}
\centerline{\tenbf EXPERIMENTAL STUDY OF TWO COMPLEX NUMBERS $\eta_{+-}$ AND
$\eta_{00}$ \footnote{Lectures
given at the Theoretical Advanced Studies Institute
in Elementary Particle Physics, University of Colorado at Boulder, Boulder,
Colorado, May29 - June24, 1994.}}
\vspace{0.8cm}
\centerline{\tenrm SUNIL V. SOMALWAR}
\baselineskip=13pt
\centerline{\tenit Department of Physics and Astronomy, Rutgers University,
P.O. Box 849, Frelinghuysen Road}
\baselineskip=12pt
\centerline{\tenit Piscataway N.J. 08855, U.S.A.}
\vspace{0.9cm}
\abstracts{Recent and upcoming Fermilab experiments to probe the origin
of CP violation and to search for CPT violation are reviewed for an
audience of theoretical particle physics graduate students.}
\vfil
\twelverm   
\baselineskip=14pt
\section{CP Violation}
I will begin with a quote from J. Cronin \cite{Cronin}, who co-discovered CP
violation with Christenson, Fitch, and Turlay in 1964 \cite{Christenson}.
``The discovery of CP violation was a complete surprise to experimentalists
(who) found it as well as to the physics community at large....The experiment
that made the discovery was not motivated by the idea that such a violation
might exist.''  They found that $K_L$ ($L$ for long-lived), thought to be
the pure CP-odd mixture of $K^0$ and $\overline{K}^0$,
decayed into the CP-even $\pi^+ \pi^-$ state at the approximate rate of
two per thousand, indicating a small CP-even contamination in the $K_L$.
This CP-violation also shows up as the charge asymmetry in the
semileptonic decays, and in the existence of other CP-even decay modes
such as $\pi^0 \pi^0$. (It is interesting to note that no
manifestation has been found outside the $K_L$ system yet - not even in
its quantum-mechanical cousin $K_S$ ($S$ for short-lived),
which is required by the CPT symmetry to
have the same proportion of wrong-CP admixture in it.)

On the theoretical side, we can not yet claim to understand the
phenomenon of CP violation. The standard
model offers the best explanation as yet through the
imaginary phase of the Cabibbo-Kobayashi-Maskawa (CKM) mixing matrix.
However, the predictions of this hypothesis have so far eluded
the vigourous experimental scrutiny described in this article.

Since CP is a very fundamental symmetry, it is not surprising that it
has implications in cosmology. Our present understanding of baryogenesis
(why matter far outweighs antimatter in the universe) requires
CP violation as a necessary ingredient.  It is possible that
the conventional CP violation, i.e., observed for
$K_L$ and presumably originating in the CKM mixing matrix, is sufficient
for this purpose \cite{Farrar}, but this subject is controversial
\cite{Gavela}. If a new source of CP violation is
necessary for baryogenesis, it would be surprising if the origins of the two
types of CP violation were totally dissimilar.

The outline of the CP section of this paper is as follows.
I will briefly review how the accomodation of CP violation in the
standard model leads to the prediction of a very small `direct'
(i.e., occuring in kaon decay as opposed to in kaon mixing)
CP violation.  Experimental
signatures of the direct CP violation can occur in a variety of decay modes
in the kaon system.  I will first describe the signature
in the 2$\pi$ decay of
the $K_L$, characterized by the ratio $\epsilon^{'} / \epsilon$. Then
I will discuss Fermilab fixed-target experiments E731 (concluded)
and E832 (expected to run in 1996).

A good reference for this subject is a recent review article on the
search for direct CP violation
by Winstein and Wolfenstein \cite{Winwolf}.  This review includes
a comprehensive compilation of references on CP violation.  In this article,
I have avoided duplicating their list of references.

\section{Standard Model and Direct CP Violation}
The minimal standard model with its three generations allows the
generation-mixing CKM mixing matrix to have one non-trivial phase.
The coupling between the quarks and the W boson
involves the CKM matrix, and the introduction of the phase in the coupling
allows
CP violating processes to occur.  This CP-violating phase appears as
a free parameter in the standard model, and its value will have to
be inferred from existing and future experimental measurements.
To this extent, the standard model incorporates CP violation on a
somewhat ad-hoc basis.
The second-order box diagram (Fig. 1) shows how the $K^0-\bar{K^0}$
mixing takes place.

%
The calculation of the mixing parameter $\epsilon$
is complicated due to QCD corrections, dependence on the quark masses,
etc.  Due to the uncertainties involved, one can not
translate exactly the measured
value of $\epsilon$ into the CP violating phase
of the CKM matrix, but $\epsilon$ provides one of the constraints.

Although this hypothesis for explaining $\epsilon$ is somewhat ad-hoc,
fortunately there is a
prediction. Once the Pandora's box in Fig. 1 is opened to explain mixing,
other diagrams (such as the one shown in Fig. 2, called
the gluonic Penguin) come marching right out.

%
Due to these diagrams, in addition to the CP violating
{\it mixing} ($\epsilon$) of CP-even kaon in the CP-odd kaon, one also
gets CP violation in the {\it decay} of the kaon, the so-called `direct'
CP violation characterized by Real($\epsilon^{'} / \epsilon$).
The calculation of $\epsilon^{'} / \epsilon$ is also plagued by
similar difficulties.  This is an evolving subject, but most calculations
indicate that Real($\epsilon^{'} / \epsilon$) is a small number in the
range of 10$^{-3}$.

\section{Observable Manifestations of Direct CP Violation}
Let us consider the CP violating decay mode $K_L \rightarrow 2\pi$.
Since the expected value of Real($\epsilon^{'} / \epsilon$) is small, the
CP violation in this decay mode is mostly ascribed to the CP-even
$\epsilon K_1$ contamination in the CP-odd state $K_2$.  The puzzle is
how to tell apart a small direct CP violating
$K_2 \rightarrow 2\pi$
decay component in the presence of a much more dominant
$K_1 \rightarrow 2\pi$ component that arises from mixing.  A brute
force approach would require both the measurement {\it and} the prediction of
$\epsilon$ accurate to better than $10^{-3}$, which is clearly unrealistic.

Fortunately, the 2$\pi$ final state comes in two varieties - $\pi^+ \pi^-$
and $2\pi^0$, each a different isospin combination. The CP violating
amplitude ratios for these states are defined as follows.
\begin{equation}
\eta_{+-} \equiv |\eta_{+-}|e^{i\phi_{+-}}
= \frac{A(K_L \rightarrow \pi^+\pi^-)}{A(K_S \rightarrow \pi^+\pi^-)}
\end{equation}
\begin{equation}
\eta_{00} \equiv |\eta_{00}|e^{i\phi_{00}}
= \frac{A(K_L \rightarrow \pi^0\pi^0)}{A(K_S \rightarrow \pi^0\pi^0)}
\end{equation}
Note that the initial and final states in the numerator and the denominator
are physical states.  Let us now define the
direct CP violating quantity $\epsilon^{'}$.
\begin{equation}
\epsilon^{'} = \frac{1}{\sqrt{2}}\frac{A(K_2 \rightarrow 2\pi, I=2)}
                                    {A(K_1 \rightarrow 2\pi, I=0)}
\end{equation}
$\epsilon^{'}$ as defined above is an abstract quantity in the sense
that both the initial
and final states in the numerator and the denominator are not physical
states but are eigenstates of CP and isospin.  Note also that the numerator
not only exhibits direct CP violation, but also breaks the $\Delta I = 1/2$
rule.  One can now use Clebsch-Gordon coefficients to go from the pure
isospin states to the physical 2$\pi$ states,
and the known CP violation quantities to express the pure CP states $K_1$
and $K_2$ in terms of the weak eigenstates $K_S$ and $K_L$.  After all the
dust settles, the abstract definition of $\epsilon^{'}$
translates into a small, but {\it experimentally accessible}
inequality of CP violation in the $\pi^+ \pi^-$ and the
$\pi^0 \pi^0$ decay modes of the $K_L$, normalized to the corresponding
CP conserving modes of the $K_S$.
\begin{equation}
{\rm Real}(\epsilon^{'} / \epsilon) \approx 1/6 \left( \left|
\frac{A(K_L \rightarrow \pi^+ \pi^-) / A(K_S \rightarrow \pi^+ \pi^-)}
     {A(K_L \rightarrow \pi^0 \pi^0) / A(K_S \rightarrow \pi^0 \pi^0)}
\right| ^2 - 1 \right)
\end{equation}
We have ignored some higher order quantities where appropriate, and also
taken into account the known suppression due to the $\Delta I = 1/2$ rule.
The same expression can also be written as
\begin{equation}
{\rm Real}(\epsilon^{'} / \epsilon) \approx 1/6 \left( \left|
\frac{\eta_{+-}}{\eta_{00}}\right| ^2 - 1 \right)
\end{equation}

\section{Search for Direct CP violation}
As can be seen from Eq. (4) above, a measurement of
Real$(\epsilon^{'} / \epsilon)$ involves four different decay modes.  The
branching ratio for the decay modes $K_L \rightarrow \pi^+ \pi^-$ is compared
to the branching ratio for $K_L \rightarrow \pi^0 \pi^0$, after normalizing
to the corresponding branching ratios for the $K_S$.
One may be tempted to look up the four branching ratios and calculate
$\epsilon^{'}$, only to find that the double ratio is known to approximately
5\% accuracy, whereas the desired precision is in the range of $10^{-3}$.
A better (but still challenging) experiment
measures the double ratio of the four
modes simultaneously, eliminating the common systematic errors.
In a nutshell, an experiment to measure Real$(\epsilon^{'} / \epsilon)$
is a counting experiment for these four decay modes.  Difficulty arises in
obtaining sufficient statistics while maintaining the precise knowledge of
the relative detection efficiencies for the four decay modes.  For example,
since the lifetimes of the $K_L$ and the
$K_S$ are very different, change in detector acceptance along the
flight path of the kaons results in different detector
acceptance for the $K_L$ and the $K_S$ decays.  This difference
in acceptance has to
be understood very well in order to correctly estimate the {\it actual}
number of decays that took place in the detector from the number
of decays {\it recorded} by the detector.

There are two experimental groups currently trying to establish the
phenomenon of direct CP violation in the $2\pi$ decay mode, one is based
at Fermilab (E731/E832),
and the other is based at CERN (NA31/NA48). Both are fixed-target
experiments, with different measurement techniques.  E731 and NA31
have concluded their analysis, and the next genertion experiments E832
and NA48 are expected to be taking data in 1996.
I will concentrate on
the Fermilab experimental technique in this article.

Fermilab E731 collected data during the 1987/88 Fermilab fixed-target
run.  Fig. 3 shows the schematic of the E731 detector.
The key feature
of the experiment is the reduction of systematic error
by using two beams simultaneously, one with mostly $K_L$ decays, and the
other with mostly $K_S$ decays. Two parallel beams were produced
by striking protons on a common target placed approximately 110 meters
upstream of the decay volume.  Given the large travel distance, almost
all the $K_S$'s decayed out and only the $K_L$'s survived in both beams.
$K_S$'s were produced by placing a two interaction
length Boron Carbide regenerator in one of the beams.  The $K_L$
and $K_S$ decays were collected simultaneously, thus greatly
reducing the systematic error due to accidental activity, rate dependent
effects, etc.  The regenerator alternated from beam to beam every
proton spill (approximately once a minute) to cancel the errors due
to the detector and beam asymmetries.  Further details
can be found elsewhere \cite{Gibbons}.

To record the $\pi^+ \pi^-$ decay modes,
the charged particle  tracks  and momenta were measured  with a spectrometer
system consisting of four  precision  drift  chambers and a large
aperture analysis magnet
placed  between the 2nd and the 3rd drift chambers. The  position and momentum
resolutions of this spectrometer were 0.1mm and (0.45 + 0.011p/GeV/c )\%,
respectively.  The charged mode ($\pi^+ \pi^-$)
trigger was essentially geometric in
nature, and made  use  of  scintillator  bank  signals and the hit
information from the two halves of one of the drift chambers.

The energies and positions of the four photons from the $2\pi^0$ decay modes
were
measured with a fine-resolution  electromagnetic calorimeter array of
804 lead-glass blocks.  The energy
resolution for electrons was (1 + 5/$\sqrt{E/GeV}$)\%.  The photon energy
resolution  had  an  extra  one  percent  constant  term  due  to  the
fluctuations in the conversion depth  of  the photon.  The gain of each
calorimeter block was tracked with a flash-lamp system throughout
the run.The trigger for
the neutral mode (2$\pi^0$)was formed by  counting the number of
clusters in the
calorimeter and requiring a significant (30 GeV) energy deposit.

The event sample consisted of 410k events of the CP violating mode
$K_L \rightarrow 2\pi^0$, normalized to the 800k
$K_S \rightarrow 2\pi^0$ events, and similarly 327k
$K_L \rightarrow \pi^+ \pi^-$ events normalized to 1060k
$K_S \rightarrow \pi^+ \pi^-$ events.  This sample, after correcting
for acceptance, decided the measured value of
Real($\epsilon^{'}/ \epsilon$).  However, to make sure that the
detector was understood properly, much higher statistics samples
of non-CP violating modes were also recorded.  The $\pi^+ \pi^-$ mode
acceptance was scrutinized using several million
$K_L \rightarrow \pi^{\pm} e^{\mp} \nu$ decays, whereas several million
$K_L \rightarrow 3\pi^0$ decays served the same function for the
2$\pi^0$ sample. These  high  statistics   decay   samples   were  used
extensively  for
calibration, aperture determination, and acceptance studies.

Once the signal events are reconstructed, one needs to know the
detector acceptance to high accuracy.  In particular, the detector
acceptance difference for $K_L$ and $K_S$ has to be understood very
well in both neutral and charged modes.
This is achieved with the aid of a detailed detector simulation, which has
a small number of adjustable parameters such as the kaon production
momentum
spectrum, proton beam targetting angle, and the kaon beam collimator
positions.  The accuracy of
the simulation was judged by juxtaposing many data
distributions against the corresponding predictions obtained from the
simulation. Fig. 4, for example,
shows the comparison of data and simulation for the decay vertex (Z) of the
CP violating $\pi^+\pi^-$ decays in the $K_L$ beam.
\noindent
The good quality of agreement between the data
and the simulation for the signal modes as well as the high-statistics
modes implies that the detector properties are well-understood, i.e.,
the acceptance is known accurately. The evaluation of
Real($\epsilon^{'}/ \epsilon$) is now straightforward.
The number of 2$\pi$ decays in the vacuum beam and the regenerator beam were
corrected for acceptance and backgrounds. The wavefunction for the coherently
regenerated kaons in the regenerator beam is given by
$|K_L> + \rho|K_S>$,
where $\rho$ is the coherent regeneration amplitude. The acceptance corrected
data is used to extract the ratios
$\rho / \eta_{+-}$, and $\rho / \eta_{00}$ of the regneration amplitude
to the CP-violating amplitudes in the charged and the neutral
mode, respectively.  The splitting between the two ratios in a combined fit
with common regeneration amplitude gives the value
of Real($\epsilon^{'}/  \epsilon$).  The value obtained from the entire
1987-88 data sample was (7.4 $\pm$ 5.2)$\times 10^{-4}$, where the error
is statistical only.

   The five major sources of systematic error were :
calibration uncertainties (photons are hard to measure!), uncertainties
associated with one of the trigger planes,accidental (rate) effects,
background subtractions, and finally acceptance corrections.  The combined
systematic error in $\epsilon^{'}/  \epsilon$
from these and some other minor sources amounted to
2.9$\times 10^{-4}$, thus making the result
(7.4 $\pm$ 5.2 $\pm$2.9)$\times 10^{-4}$.

\section{Direct CP Violation - Status and Future}

The E731 result,
Real($\epsilon^{'}/ \epsilon$)=(7.4 $\pm$ 6.0)$\times 10^{-4}$, is
consistent with the null result.  The CERN group (NA31) \cite{Barr}
gets a value
of (23 $\pm$ 3.6 $\pm$ 5.4)$\times 10^{-4}$ from their 1986,1988,and 1989
runs.  The NA31 value is approximately 3.5 standard deviation away from
zero.  Given the disagreement between the two results,
the errors have to be inflated artificially to achieve unit
$\chi^2$ for the combined result.  This procedure gives
Real($\epsilon^{'}/ \epsilon$)=(14 $\pm$ 8)$\times 10^{-4}$, which
is not significant enough to claim the discovery of direct CP violation
in the 2$\pi$ modes.

Both the Fermilab and CERN groups are currently building significantly
improved experiments, with the goal of getting the errors down to
1-2$\times 10^{-4}$. The Fermilab E832 group will use essentially the same
technique as in E731.  It will have an improved beamline to cleanly deliver
more kaons, and a detector with improved resolution that will also
be able to handle higher rates. In particular,
replacing the Lead-glass calorimeter array by a Cesium Iodide scintillator
crystal array will significantly improve the uncertainties associated with
the measurement of photons.  The CERN NA48
group is planning to use a calorimeter
with liquid krypton.  Both the calorimeters are expected to withstand the
higher rate needed to accumulate greater statistics.

If CP violation is indeed explained by the standard model, the direct
CP violation effect in the 2$\pi$ mode could be established with a
measurement precision of 1-2$\times 10^{-4}$ over the next few years.

Direct CP violation is also expected to be seen in other decay modes of
the $K_L$, such as $\pi^0 e^+ e^-$ and $\pi^0 \nu \bar{\nu}$.  In fact,
``$\epsilon^{'} / \epsilon$'' for these modes is expected to be comparable
to unity or even higher.  However, these decay modes are very rare and
experimentally challenging.  There is ongoing experimental effort (Fermilab
E799) to search for these and other rare decay modes.

\section{CPT Symmetry}
While CP symmetry is a fundamental symmetry with implications for
cosmology, CPT symmetry is even more fundamental  - it is difficult to
construct a field
theory without the CPT symmetry.  All local quantum field theories obey
CPT symmetry \cite{Luders,Pauli}, with minimal general assumptions
such as Lorentz invariance and asympotically free states.  At the same time,
P and CP symmetries have
been experimentally dismantled over time, and it is fair to ask whether the
CPT symmetry would also be vulnerable to sensitive probes.
It is amusing to
note that the CPT theorem was published
contemporaneously with the famous paper by Lee and Yang \cite{Leeyang}
bringing down the P symmetry.  Conceptually, one might wonder about
the status of CPT when the
interactions are not local (as in superstring theories), or
when the states are not asymptotically free (as in Quantum
Chromodynamics).  There is also the possibility of CPT symmetry being
broken in the as yet unformulated quantum theory of gravity, or by some
non-quantum mechanical effects.

The question of possible CPT violation
takes on special relevance for the neutral kaon system, which is unique
in having
exhibited CP violation and also in offering us a remarkably sensitive
interferometer with
precision approaching that of the Planck scale. The ratio of the difference of
$K_L$ and $K_S$ mass to the $K_L$ mass is approximately 7 x 10$^{-15}$,
which is a phenomenally small number.

A good reference on CPT symmetry in the neutral kaon
system is a paper by Barmin \cite{Barmin} and colleagues.  Although the
experimental status has changed significantly since the publication of this
article, it is still a good reference for phenomenology in the neutral
kaon sector.

\section{Search for CPT Violation}
CPT symmetry manifests itself in the form of equality of particle-antiparticle
masses, lifetimes (or equivalently, total decay rates), moments, etc.
For the neutral kaon \cite{Barmin}, CPT symmetry places
constraints on the values of the phases of CP violating amplitude ratios
$\eta_{+-}$ and $\eta_{00}$.
Two significant experimental CPT tests have been recently done in the
kaon sector. Recall that direct CP violation causes the
{\it magnitudes} of the amplitude ratios
$\eta_{+-}$ and $\eta_{00}$ to split (see Eq. (5)).
Similarly, CPT violation would cause the {\it phases} $\phi_{+-}$ and
$\phi_{00}$ of these amplitude ratios to split.  CPT symmetry requires
that the phase difference $\Delta \phi \equiv \phi_{00} - \phi_{+-}$ be
significantly less than 1$^0$.  A second consequence of CPT symmetry
is that these phases should also be also very close to the so-called
superweak phase
$\phi_{sw} \equiv {\rm tan}^{-1}(2\Delta m/ \Delta \Gamma)$, where
$\Delta m$ is the $K_L - K_S$ mass difference, and $\Delta \Gamma$ is the
$K_S - K_L$ decay width difference $\Gamma_S - \Gamma_L$.

The $\pi^+ \pi^-$ decays from a coherently regenerated kaon beam exhibit
an interference pattern due to the non-zero mass
difference $\Delta m$, as indicated in the following rate equation.
\begin{equation}
{\rm Rate}(\pi^+ \pi^-)=|\eta_{+-}|^2 e^{-\Gamma_{L}t}
                  + |\rho|^2 e^{-\Gamma_{S}t}
+ 2|\rho\eta_{+-}|e^{-\frac{\Gamma_S + \Gamma_L}{2}t}
{\rm cos}(\Delta mt + \phi_{\rho} - \phi_{+-})
\end{equation}
where $\rho \equiv |\rho|e^{i\phi_{\rho}}$ is the regeneration amplitude.
A similar interference is also seen in the 2$\pi^0$ decay mode.
\begin{equation}
{\rm Rate}(\pi^0 \pi^0)=|\eta_{00}|^2 e^{-\Gamma_{L}t}
                  + |\rho|^2 e^{-\Gamma_{S}t}
+ 2|\rho\eta_{00}|e^{-\frac{\Gamma_S + \Gamma_L}{2}t}
{\rm cos}(\Delta mt + \phi_{\rho} - \phi_{00})
\end{equation}
The argument of the cosine term in the two equations above shows that
the phase difference $\Delta \phi$ can be measured experimentally
by looking for a relative shift in the interference patterns for the
$\pi^+ \pi^-$ and the 2$\pi^0$ modes.  The equality of $\phi_{+-}$ and
$\phi_{sw}$, which is the second CPT test, can also experimentally tested
by extracting $\phi_{\rho} - \phi_{+-}$ from the $\pi^+ \pi^-$
interferenece pattern, and then using the value of $\phi_{\rho}$ obtained
using analyticity assumption \cite{Gibbons2}.

Both Fermilab E731 and its successor experiment E773 have performed these
two CPT tests. Fig. 5 shows an example
of an extracted interference curve from E773.
E731 reported
\cite{Gibbons2} $\Delta \phi = -1.6^0 \pm 1.2^0$, which is consistent
with zero, and $\phi_{+-} = 42.2^0 \pm 1.4^0$, which is consistent
with $\phi_{sw} = 43.4^0$, also measured by E731. The preliminary
values reported by E773 \cite{APS} also indicate no evidence for CPT violation.
The final results from E773 are expected to place CPT limits
with $1^0$ error on $\Delta \phi$, and $0.8^0$ error on $\phi_{+-}$.
The CPT symmetry appears to be safe for the time being, but data from
the next round of $\epsilon{'}/ \epsilon$ experiments will probe the CPT
symmetry to a even greater accuracy.

\section{Acknowledgements}
I would like to thank my collaborators from Fermilab experiments
E731, E773, E799, and E832.  I would also like to thank J.F.Donoghue
and K.T. Mahanthappa for their role in organizing TASI'94, and G. Thomson
for comments on the manuscript.
This work was supported by the NSF grant PHY 90-19706, NSF Young
Investigator award PHY-93-57189, and by equipment support from
the Hamamatsu Corporation (Japan).


\end{document}